\input jnl
\input reforder
\input eqnorder
\def\3he{{$^3${\rm He}}}

\def\eg{{\it e.g.,\ }}
\def\ie{{\it i.e.,\ }}
\def\etal{{\it et al.}}
\def\etc{{\it etc.}}

\def\slD{\raise.15ex\hbox{$/$}\kern-.53em\hbox{$D$}}
\def\dsl{\raise.15ex\hbox{$/$}\kern-.57em\hbox{$\Delta$}}
\def\slp{{\raise.15ex\hbox{$/$}\kern-.57em\hbox{$\partial$}}}
\def\nsl{\raise.15ex\hbox{$/$}\kern-.57em\hbox{$\nabla$}}
\def\sla{\raise.15ex\hbox{$/$}\kern-.57em\hbox{$\rightarrow$}}
\def\slla{\raise.15ex\hbox{$/$}\kern-.57em\hbox{$\lambda$}}
\def\slb{\raise.15ex\hbox{$/$}\kern-.57em\hbox{$b$}}
\def\lnp{\raise.15ex\hbox{$/$}\kern-.57em\hbox{$p$}}
\def\lnk{\raise.15ex\hbox{$/$}\kern-.57em\hbox{$k$}}
\def\lnK{\raise.15ex\hbox{$/$}\kern-.57em\hbox{$K$}}
\def\lnq{\raise.15ex\hbox{$/$}\kern-.57em\hbox{$q$}}

\def\psib{{\overline\psi}}

\def\cL{{\cal L}}
\def\cM{{\cal M}}


\def\pmb#1{\setbox0=\hbox{$#1$}%
\kern-.025em\copy0\kern-\wd0
\kern.05em\copy0\kern-\wd0
\kern-.025em\raise.0433em\box0 }

\def\q2{{Q^2}}
\def\gtwid{\raise.3ex\hbox{$>$\kern-.75em\lower1ex\hbox{$\sim$}}}
\def\ltwid{\raise.3ex\hbox{$<$\kern-.75em\lower1ex\hbox{$\sim$}}}
\def\12{{1\over2}}
\def\part{\partial}

\def\low#1{\lower.5ex\hbox{${}_#1$}}

\def\psl{\raise.15ex\hbox{$/$}\kern-.57em\hbox{$\partial$}}
\def\partt{\raise.15ex\hbox{$\widetilde$}{\kern-.37em\hbox{$\partial$}}}

\def\topppageno1{\global\footline={\hfil}\global\headline
={\ifnum\pageno<\firstpageno{\hfil}\else{\hss\twelverm --\ \folio
\ --\hss}\fi}}
 
\def\toppageno2{\global\footline={\hfil}\global\headline
={\ifnum\pageno<\firstpageno{\hfil}\else{\rightline{\hfill\hfill
\twelverm \ \folio
\ \hss}}\fi}}

\def\prl#1{Phys.\ Rev.\ Lett.\ {\bf #1}}
\def\prd#1{Phys.\ Rev.\ {\bf D#1}}
\def\plb#1{Phys.\ Lett.\ {\bf #1B}}
\def\npb#1{Nucl.\ Phys.\ {\bf B#1}}

\def\ie{{\it i.e.},\ }
\def\eg{{\it e.g.},\ }

\def\ref#1{${}^{#1}$}
\def\nsection#1 #2{\leftline{\rlap{#1}\indent\relax #2}}

\def\prl#1{Phys.\ Rev.\ Lett.\ {\bf #1}}
\def\prd#1{Phys.\ Rev.\ {\bf D#1}}
\def\plb#1{Phys.\ Lett.\ {\bf #1B}}
\def\npb#1{Nucl.\ Phys.\ {\bf B#1}}

\input epsf.tex

\def\qa{{quenched approximation}}

\def\psib{{\overline{\psi}}}

\def\etat{{\tilde{\eta}}}
\def\Dsl{D\!\!\!\!/\,\,}
\def\eq#1{{eq. \(#1)}}

\def\cL{{\cal L}}
\def\etc{{\it etc}}
\def\ie{{\it i.e.}}
\def\eg{{\it e.g.}}
\def\qt{{\tilde q}}
\def\qbar{{\overline q}}
\def\qtbar{{\overline\qt}}
\def\phit{{\tilde\phi}}
\def\half{{{1\over 2}}}
\def\cf{{\it cf.}}
\def\mbar{{\overline m}} 

{\parindent=0pt {April 1994, \hfill{Wash. U. HEP/94-62} }

\title How good is the quenched approximation of QCD?$\null^*$
\footnote{$\null$}{$\null^*$ Talk given at WHEPP III, Madras, India}
\author Maarten F.L. Golterman                       
\affil Department of Physics 
       Washington University 
       St. Louis, MO 63130, USA 

\abstract {The \qa\ for QCD is, at present and in the foreseeable
future, unavoidable in lattice calculations with realistic choices of
the lattice spacing, volume and quark masses. In this talk, I review an
analytic study of the effects of quenching based on chiral perturbation
theory. Quenched chiral perturbation theory leads to quantitative
insight on the difference between quenched and unquenched QCD, and
reveals clearly some of the diseases which are expected to plague
quenched QCD.} 
\endtopmatter

\subhead{\bf 1. Introduction}

The lattice formulation of QCD has proven to be a powerful tool for
computing QCD quantities of direct phenomenological interest, such as
hadron masses, decay constants, weak matrix elements, the strong
coupling constant, \etc. (For reviews see for instance refs.
[\cite{sharpe,mackenzie}], or the proceedings of Lattice 93
[\cite{lat93}].)
 
In order to perform such computations numerically, one obviously needs 
to consider a system with a finite number of degrees of freedom, which 
is accomplished by putting lattice QCD in a finite box.  This box is 
then hopefully large enough to accomodate the physics one is interested 
in without serious finite size effects.  This leads to the requirement 
that the Compton wavelength of the particles of interest is sufficiently 
smaller than the linear dimension of the box, \ie\ the mass has to be 
large enough for the particles to fit in the box. 

In order to have a small enough lattice spacing, small enough masses (in
particular for the pion) and a large enough box size, one needs a large
number of degrees of freedom in a numerical computation. It turns out
that for QCD with realistic choices of the lattice spacing $a$, volume
$V$ and the quark masses (in particular the light quark masses), the
presently available computational power is not adequate. The most severe
problem comes from the fermion determinant, the logarithm of which is a
very nonlocal part of the gluon effective action (specially for light
quark masses). This nonlocality slows down the Monte Carlo algorithms
dramatically.

In order to circumvent this problem, most numerical computations in
lattice QCD have been done in the \qa, in which one simply sets the
fermion determinant equal to one [\cite{quenched}]. This amounts to
ignoring all fermion loops which occur in QCD correlation functions
(except those put in by hand through the choice of operators on the
external lines). While some handwaving arguments exist as to why this
might not be unreasonable, the \qa\ does introduce an uncontrolled
systematic error. Since the effect of a fermion loop is roughly
inversely proportional to its mass, this error is expected to be
particularly severe for quantities involving light quarks. It appears
therefore that chiral perturbation theory ChPT maybe a useful tool for
investigating the difference between quenched and unquenched (``full")
QCD.

In this talk, I will review a systematic approach to the study of the
\qa\ through ChPT [\cite{sharpe1,us1, sharpe2,uslat92}]. There are two
reasons why ChPT is useful in this context:

$\bullet$ It turns out that ChPT can be systematically adapted to
describe the low energy sector of quenched QCD [\cite{us1}]. It will
therefore give us nontrivial, quantitative information on the difference
between quenched and full QCD.

$\bullet$ ChPT describes the approach to the chiral limit, and can be
used for extrapolation of numerical results to small masses and large
volumes. If these results come from quenched computations, one will of
course need a quenched version of ChPT. (For finite volume ChPT, see
refs. [\cite{hasleu}]. For quenched finite volume results, see refs.
[\cite{us1,sharpe2}].)

In this review, I will concentrate on the first point. I will first show
how ChPT is developed for the \qa, and then use it for a quantitative
comparison between full and quenched QCD. The quantities that I will
discuss are $f_K/f_\pi$ [\cite{us1,uslat92}] and the octet baryon masses
[\cite{labsha}].

I will then address a number of theoretical problems that arise as a 
consequence of quenching.  That such problems arise is no surprise, 
as quenching QCD mutilates the theory quite severely.  It is however 
quite instructive to see what the detailed consequences are.  

\subhead{\bf 2. Systematic ChPT for quenched QCD}

In this section I will outline the construction of a chiral effective
action for the Goldstone boson sector of quenched QCD [\cite{us1}]. I
will first introduce the formalism, and then show how it works in some
examples. For early ideas on quenched ChPT, see ref. 
[\cite{morel,sharpe1}].

We will start from a lagrangian definition of euclidean quenched QCD. 
(We will restrict ourselves entirely to the euclidean theory which can 
be defined by a pathintegral.  Hamiltonian quenched QCD presumably does 
not exist.)  To
the usual QCD lagrangian with three flavors of quarks $q_a$, $a=u,d,s$,
we add three ghost quarks $\qt_a$ with exactly the same quantum numbers
and masses $m_a$, but with opposite, bosonic, statistics [\cite{morel}]: 
$$\cL_{\rm quarks}=
\sum_a\qbar_a(\Dsl+m_a)q_a+\sum_a\qtbar_a(\Dsl+m_a)\qt_a,\eqno(qqcd)$$ 
where $\Dsl$ is the covariant derivative coupling the quark and ghost
quarks to the gluon field. The gluon effective action produced by
integrating over the quark- and ghost quarkfields vanishes, since the
fermion determinant of the quark sector is exactly cancelled by that of
the ghost sector.  Note that the ghost quarks violate the 
spin-statistics theorem.

We will now assume that mesons are formed as (ghost) quark - (ghost)
antiquark pairs just like in ordinary QCD. This is basically equivalent
to the notion that it is the dynamics of the gluons which leads to
confinement and chiral symmetry breaking. The Goldstone particle
spectrum of quenched QCD will then contain not only $q\qbar$, but also
$\qt\qtbar$, $q\qtbar$ and $\qt\qbar$ bound states.  We will denote this 
36-plet by
$$\Phi\equiv\left(\matrix{\phi&\chi^\dagger\cr
\chi&\phit\cr}\right)\sim
\left(\matrix{q\qbar&q\qtbar\cr\qt\qbar&\qt\qtbar\cr}\right).\eqno(octet)$$
Note that the fields $\chi$ and $\chi^\dagger$ describe Goldstone 
fermions.

The quenched QCD lagrangian \eq{qqcd} for vanishing quark masses has a
much larger symmetry group than the usual $U(3)_L\times U(3)_R$ flavor
group; it is invariant under the graded group $U(3|3)_L\times U(3|3)_R$
[\cite{us1}],
where $U(3|3)$ is a graded version of $U(6)$ since it mixes the fermion
and boson fields $q$ and $\qt$. Writing an element $U$ of $U(3|3)$ in
block form as
$$U=\left(\matrix{A&C\cr D&B\cr}\right),\eqno(element)$$
the $3\times 3$ matrices $A$ and $B$ consist of commuting numbers, while 
the $3\times 3$ matrices $C$ and $D$ consist of anticommuting numbers.

We can now construct a low energy effective action for the Goldstone 
modes along the usual lines.  We introduce the unitary field
$$\Sigma=exp(2i\Phi/f),\eqno(sigma)$$
which transforms as $\Sigma\to U_L\Sigma U_R^\dagger$ with $U_L$ and 
$U_R$ elements of $U(3|3)$.  Because we are dealing here with a graded 
group, in order to build invariants, we need to use the supertrace $str$ 
and the superdeterminant $sdet$ instead of the normal trace and 
determinant, with [\cite{dewitt}]
$$\eqalignno{
str(U)&=tr(A)-tr(B),\cr
sdet(U)&=exp(str\log{(U)})=det(A-CB^{-1}D)/det(B).&(strsdet)
}$$
To lowest order in the derivative expansion, and to lowest order in the 
quark masses, the chiral effective lagrangian consistent with our graded 
symmetry group is
$$\cL_0={{f^2}\over 8}str(\partial_\mu\Sigma\partial_\mu\Sigma^\dagger)
-v\;str(\cM\Sigma+\cM\Sigma^\dagger),\eqno(clo)$$
where $\cM$ is the quark mass matrix
$$\cM=\left(\matrix{M&0\cr 0&M\cr}\right),\ \ \ \ \ \ \ \ 
M=\left(\matrix{m_u&0&0\cr 0&m_d&0\cr 
0&0&m_s\cr}\right).\eqno(massmatrix)$$
$f$ and $v$ are bare coupling constants which are not yet determined at 
this stage.

The symmetry group is broken by the anomaly to the smaller group\break
$[SU(3|3)_L\times SU(3|3)_R]{\bigcirc\kern -0.30cm s\;}U(1)$ (the semidirect
product arises as a consequence of the graded nature of the groups
involved; the details are irrelevant for this talk).  $SU(3|3)$ consists 
of all elements $U\in U(3|3)$ with $sdet(U)=1$.  The anomalous field
is $\Phi_0=(\eta'-\etat')/\sqrt{2}$, where the relative minus sign comes
from the fact that in order to get a nonvanishing triangle diagram, one
needs to choose opposite explicit signs for the quark and ghost quark
loops, due to the different statistics of these fields.  $\eta'$ is the 
field describing the normal $\eta'$ particle, while $\etat'$ is the 
ghost $\eta'$ consisting of ghost quarks and ghost antiquarks.
We will call the field $\Phi_0$ the super-$\eta'$ field.  The field 
$\Phi_0\propto str\log{\Sigma}=\log{sdet\;\Sigma}$ is invariant under 
the smaller symmetry group, and we should include arbitrary functions of 
this field in our effective lagrangian.  Following ref. [\cite{gasleu}], 
the correct chiral effective lagrangian is
$$\eqalignno{
\cL=&V_1(\Phi_0)str(\partial_\mu\Sigma\partial_\mu\Sigma^\dagger)
-V_2(\Phi_0)str(\cM\Sigma+\cM\Sigma^\dagger)\cr
&+V_0(\Phi_0)+V_5(\Phi_0)(\partial_\mu\Phi_0)^2,&(chptlag)
}$$
where the function multiplying $i\;str(\cM\Sigma-\cM\Sigma^\dagger)$ can
be chosen equal to zero after a field redefinition. This lagrangian
describes quenched ChPT systematically, as we will show now with a few
examples.

For our first example, let us isolate just the quadratic terms for 
the fields $\eta'$ and $\etat'$, choosing degenerate quark masses for 
simplicity.  We expand
$$\eqalignno{
V_1(\Phi_0)&={{f^2}\over 8}+\dots,\cr
V_2(\Phi_0)&=v+\dots,\cr
V_0(\Phi_0)&={\rm constant}+\mu^2\Phi_0^2+\dots,&(expand)\cr
V_5(\Phi_0)&=\alpha+\dots,
}$$
and obtain
$$\eqalignno{
\cL(\eta',\etat')=&\half(\partial_\mu\eta')^2
-\half(\partial_\mu\etat')^2+\half\alpha(\partial_\mu\eta'
-\partial_\mu\etat')^2\cr
&+\half m_\pi^2(\eta')^2-\half m_\pi^2(\etat')^2+\half\mu^2(\eta'-\etat')^2
+\dots,&(etaplag)
}$$
where $m_\pi^2=8mv/f^2$. The relative minus signs between the $\eta'$ and
$\etat'$ terms in \eq{etaplag} come from the supertraces in
\eq{chptlag}, and are related to the graded nature of the chiral
symmetry group of quenched QCD.  

The inverse propagator in momentum space,
$$(p^2+m_\pi^2)\left(\matrix{1&0\cr 0&-1}\right)
+(\mu^2+\alpha p^2)\left(\matrix{1&-1\cr -1&1}\right),\eqno(invprop)$$ 
clearly cannot be diagonalized in a $p$ independent way, which is quite
different from what one would expect from a normal field theory!
Treating the $\mu^2+\alpha p^2$ 
term as a twopoint vertex, one can easily show
that the repetition of this vertex on one meson line vanishes, due to
the fact that the propagator matrix $\left(\matrix{1&0\cr 0&-1}\right)$
multiplied on both sides by the vertex matrix $\left(\matrix{1&-1\cr
-1&1}\right)$ gives zero. This result coincides exactly
with what one would expect from the quark flow picture for $\eta'$
propagation, as depicted in fig. 1. The straight-through and double
hairpin contributions do not contain any virtual quark loops, and are
therefore present in the \qa. All other contributions should vanish
because they do contain virtual quark loops, and this is exactly what
happens as a consequence of the (admittedly strange) Feynman rules for
the propagator in the $\eta'$--$\etat'$ sector!  This propagator is 
given by the inverse of \eq{invprop} and reads
$${1\over{p^2+m_\pi^2}}\left(\matrix{1&0\cr 0&-1}\right)
-{{\mu^2+\alpha p^2}\over{(p^2+m_\pi^2)^2}}\left(\matrix{1&1\cr 1&1}\right),
\eqno(prop)$$
in which the two terms correspond to the two first diagrams in fig. 1.

\vskip 3cm\par
\centerline{
Figure 1. {\it Quark flow diagrams for the $\eta'$ propagator in full 
QCD.}}
\par \medskip

From \eq{prop} we learn several things. First, because $\mu^2$, which in
full ChPT would correspond to the singlet part of the $\eta'$ mass,
appears in the numerator, we need to keep the $\eta'$ (and its ghost
partner) in quenched ChPT: it cannot be decoupled by taking $\mu^2$
large. Second, this ``propagator" is definitely sick, due to the double
pole term. It should be stressed here that this double pole term is an
unescapable consequence of quenched QCD, and does not result from our
way of setting up chiral perturbation theory. In the case of
nondegenerate quark masses, this double pole also shows up in
the $\pi^0$ and $\eta$ propagators, due to mixing with the $\eta'$. I will
return to these strange properties of quenched QCD in section 4.

\vbox{
\vskip 3cm\par
\centerline{
Figure 2. {\it One loop pion selfenergy in quenched ChPT.}}
}
\par \medskip

As a second example, we will consider the (charged) pion selfenergy at 
one loop, again with degenerate quark masses.  I will set $\alpha=0$ for 
simplicity.  At one loop, the pion selfenergy only contains tadpoles, 
with either $\phi$ or $\chi$ lines (\cf\ \eq{octet}) on the loop.  Also, 
on the $\phi$ loop, one can have an arbitrary number of insertions of 
the vertex $\mu^2$ if the internal $\phi$ line is an $SU(3)$ singlet.  
These various contributions are drawn in fig. 2, where a solid line 
denotes a $\phi$ line, a dashed line denotes a $\chi$ line, and a cross
denotes a $\mu^2$ vertex.  One finds that the diagrams with the $\phi$ 
and $\chi$ lines on the loop without any crosses cancel, and then, of 
course, that the diagrams with more than one cross vanish, using our 
earlier result for the $\eta'$--$\etat'$ propagator.  We are left with 
only one term, and the result is
$$\Sigma_\pi(p)={{2m_\pi^2}\over{3f^2}}\mu^2
\int {{d^4k}\over{(2\pi)^4}}{1\over{(k^2+m_\pi^2)^2}}.\eqno(piself)$$
The pion selfenergy is logarithmically divergent, but the origin of this 
divergence is completely different from those that arise in the 
unquenched theory, as it is proportional to $\mu^2$.  One can easily 
convince oneself that the diagrams in fig. 2 which cancel or vanish 
correspond to diagrams with virtual quark or ghost quark loops in the 
quark flow picture.  (For early discussions of the quenched pion 
selfenergy in the quark flow picture, see refs. [\cite{morel,sharpe1}].)

Before I go on to look at some quantitative results, I would like to
discuss one aspect of the chiral expansion in quenched ChPT. The chiral
expansion is basically an expansion in the pion mass (see \eg\ ref.
[\cite{weinberg}]). However, as we have argued above, in quenched ChPT
there is unavoidably another mass scale, namely the singlet part of the
$\eta'$ mass, $\mu^2$. For our expansion to be systematic as an
expansion in the pion mass, we would have to sum up all orders in
$\mu^2$, at a fixed order in the pion mass. This is clearly a formidable
task. In order to avoid this complication in a systematic way, we can
think of $\mu^2/3$ (which turns out to be the natural parameter as it
appears in the chiral expansion) as an independent small parameter.
To check whether this makes any sense, one may note that from the
experimental value of the $\eta'$ mass one obtains a value
$\mu^2/3\approx (500\;MeV)^2$, which is roughly equal to the kaon mass
squared, $m_K^2$. Of course, for quenched QCD the parameter $\mu^2$ need
not have the same value, after all quenched QCD is a different
theory. A lattice computation of this parameter
[\cite{kuretal}] gives $\mu_{\rm quenched}^2/\mu_{\rm full}^2\approx
0.75$. ($\alpha$ can be estimated from $\eta$--$\eta'$ mixing, and is
very small.) Finally, one may also note that both $\mu^2$ and $\alpha$
are of order $1/N_c$, where $N_c$ is the number of colors [\cite{witven}].
I will return to this point in section 4.

\subhead{\bf 3. Quantitative comparison of quenched and full ChPT}

Let us first consider the quenched result for the ratio of the kaon and
pion decay constants $f_K$ and $f_\pi$ [\cite{us1,uslat92}]. I will set
$\alpha=0$ and take $m_u=m_d\equiv m$:
$$\left({{f_K}\over{f_\pi}}\right)^{\rm 1-loop}_{\rm quenched}=
1+{{\mu^2/3}\over{16\pi^2f_\pi^2}}
\left[{{m_K^2}\over{2(m_K^2-m_\pi^2)}}\log{\left({{2m_K^2}
\over{m_\pi^2}}-1\right)}-1\right]+(m_s-m)L.\eqno(fkfpi)$$
$L$ is a certain combination of ``low energy constants" [\cite{gasleu}].  
Since this constant is a bare parameter of the quenched chiral 
lagrangian, the result \eq{fkfpi} is not directly comparable to the 
equivalent result for the full theory.  In other words, in order to 
compare quenched and full QCD, we have to consider 
quantities which are independent of the bare 
parameters of the effective action.  (Alternatively, we would need to 
extract the values of the bare parameters from some independent 
measurement or lattice computation, in this case, we would need 
independent determinations of $L$ in both quenched and full
QCD.)  In the full theory, 
$(f_\eta f_\pi^{1/3})/f_K^{4/3}$ is such a quantity [\cite{gasleu}], but 
in the quenched theory this quantity is not well defined, due to the 
double poles which occur in the propagators of neutral mesons.  

We will therefore choose to consider a slightly different theory, in
which sufficiently many decay constants referring only to charged 
(\ie\ off-diagonal) mesons are present [\cite{uslat92}]. 
This theory is a theory
with two light quarks $m_u=m_d=m$ and two heavy quarks $m_s=m_{s'}=m'$.
This theory contains a $u\overline d$ pion $\pi$, an $s'\overline s$ 
pion $\pi'$ and a $u\overline s$
kaon $K$, with mass relation 
$$m_K^2=\half(m_\pi^2+m_{\pi'}^2).\eqno(massrel)$$
One can show that the ratio $f_K/\sqrt{f_\pi f_{\pi'}}$ is independent
of the low energy constants $L$. For the quenched theory we find 
$$\left({{f_K}\over{\sqrt{f_\pi f_{\pi'}}}}\right)^{\rm 1-loop}_{\rm 
quenched}=
1+{{\mu^2/3}\over{16\pi^2f_\pi^2}}
\left[{{m_\pi^2+m_{\pi'}^2}\over{2(m_{\pi'}^2-m_\pi^2)}}
\log{\left({{m_{\pi'}^2}
\over{m_\pi^2}}\right)}-1\right],\eqno(qratio)$$
whereas in the full theory
$$\left({{f_K}\over{\sqrt{f_\pi f_{\pi'}}}}\right)^{\rm 1-loop}_{\rm
full}=1-{1\over{64\pi^2f_\pi^2}}\left[m_\pi^2\log{\left(
{{m_K^2}\over{m_\pi^2}}\right)}+m_{\pi'}^2\log{\left(
{{m_K^2}\over{m_{\pi'}^2}}\right)}\right].\eqno(fratio)$$
Note again that the logarithms in the quenched and unquenched 
expressions are completely different in origin.  

We may now compare these two expressions using ``real world" data, where 
we'll determine the value of the $\pi'$ mass from the mass relation 
\eq{massrel}.  With $m_\pi=140\;MeV$, $m_K=494\;MeV$ and 
$\mu^2/3=0.75\times(500\;MeV)^2$ we find
$$\eqalignno{
\left({{f_K}\over{\sqrt{f_\pi f_{\pi'}}}}\right)^{\rm 1-loop}_{\rm
quenched}&=1.049,\cr
\left({{f_K}\over{\sqrt{f_\pi f_{\pi'}}}}\right)^{\rm 1-loop}_{\rm
full}&=1.023,&(results)
}$$
a difference of $3\%$.  If we choose $\mu^2/3=(500\;MeV)^2$, we find 
a difference of about $4\%$.  This difference is small.  Note however, 
that this is due to the fact that for this particular ratio, ChPT seems 
to work very well, both for the full and the quenched theories.  If one 
only considers the size of the one loop corrections (the numbers behind 
the decimal point), the quenched and full results are very different.
It is also possible, and in fact not unlikely, that part of the 
difference between the full and quenched theory gets ``washed out" by 
the fact that we are considering a ``ratio of ratios".  It follows that 
the relative difference is a lower bound on the difference between the 
quenched and full values of the decay constants.  For another quantity 
for which the difference between quenched and full ChPT has been 
calculated, see ref. [\cite{uslat92}].

Next, I will review some very recent work on baryons in quenched ChPT 
by Labrenz and Sharpe [\cite{labsha}].  They calculated the one loop 
corrections to the octet baryon mass coming from the cloud of Goldstone 
mesons.  They employed an effective lagrangian for quenched heavy baryon 
ChPT, constructed using the same techniques as described in section 2. 
In the case of degenerate quark masses, the result for the nucleon mass 
is
$$\eqalignno{
m_N=&\mbar-{{3\pi}\over 
2}(D-3F)^2{{\mu^2/3}\over{8\pi^2f_\pi^2}}m_\pi
+2(b_D-3b_F)m_\pi^2\cr
&+\left[{2\over 3}(D-3F)(2D+3\gamma)+{5\over 6}\alpha(D-3F)^2\right]
{{m_\pi^3}\over{8\pi f_\pi^2}}.&(nucleon)
}$$
In this equation, $\mbar$, $D$, $F$, $b_D$, $b_F$ and $\gamma$ are bare 
parameters which occur in the baryon effective action.  $\mbar$ is the 
bare ``average" octet mass, $D$ and $F$ are the well known baryon-meson 
couplings, $b_D$ and $b_F$ are low energy constant which arise as a 
consequence of renormalization (see for instance refs. [\cite{manjen,
berkaimei}]).  $\gamma$ is a new coupling which occurs because of the 
unavoidable presence of the super-$\eta'$ in the \qa.  
The term proportional to 
$\mu^2$ comes from a diagram with a cross on the $\phi$ internal line, 
\ie\ an insertion of the $\mu^2$ twopoint vertex.  Note that in this 
case there are 
also one loop corrections not involving $\mu^2$ which survive 
the \qa, in contrast to the pion selfenergy, \eq{piself}, or 
$f_K/f_\pi$, \eq{fkfpi}.  The authors of ref. 
[\cite{labsha}] then calculated the coefficients using full 
QCD values for the various parameters (from ref. [\cite{berkaimei}]).  
With $\alpha=0$ and $\gamma=0$ ($\gamma=0$ is consistent with available
information, which however is limited [\cite{hatsuda}]), 
they obtained
$$m_N=0.97-0.5{\delta\over{0.2}}m_\pi+3.4m_\pi^2-1.5m_\pi^3,
\eqno(bmass)$$
with $\delta\equiv\mu^2/(24\pi^2f_\pi^2)$ and $\delta\approx 0.2$ for 
the full theory.

In ref. [\cite{labsha}], \eq{nucleon} was also compared to recent numerical 
results from ref. [\cite{weinetal}].  These data are presented in fig. 3, 
where the scale $a^{-1}=1.63\;GeV$ is set by $f_\pi$ [\cite{labsha}].  
If one calculates the coefficients in \eq{nucleon} by ``fitting" the 
four data points, one finds
$$m_N=0.96-1.0m_\pi+3.6m_\pi^2-2.0m_\pi^3.\eqno(fit)$$
This is only four data points for four parameters, and the ``fit" is quite
sensitive to for instance an additional $m_\pi^4$ term.  
From the agreement between \eq{bmass} and \eq{fit} it appears that it is 
reasonable to apply ChPT to the results of ref. [\cite{weinetal}].
Note that the individual terms in \eq{bmass} are quite large for the two 
higher pion masses in fig. 3 (this is not unlike the case of unquenched 
ChPT).  From fig. 3 it is also clear that $(\mbar/f_\pi)_{\rm 
quenched}\ne (\mbar/f_\pi)_{\rm full}$ because of the term linear in 
\eq{nucleon}, which is absent in full ChPT.  

Labrenz and Sharpe then went on to consider octet mass splittings.
In order to remove 
effects which can be accomodated by a change of scale, they calculated 
the ratios
$$R_{ij}={{m_i}\over{m_j}},\ \ \ i,j=N,\Lambda,\Sigma,\Xi\eqno(R)$$
in quenched ChPT, and compared these with similar ratios obtained from 
ref. [\cite{berkaimei}].  They assumed that all bare parameters in the 
equations for the octet masses (for explicit expressions, see their 
paper) are equal in the full and quenched theory, and then calculated 
the ratios
$$r_{ij}={{R_{ij}^{\rm quenched}}\over{R_{ij}^{\rm full}}}.\eqno(r)$$
With the assumption that the bare parameters of the quenched and full 
theories are equal, $b_D$ and $b_F$ drop out of the ratios, and with 
$\gamma=0$, $\alpha=0$ and $D$ and $F$ equal to their full QCD values, 
they obtain
$$\eqalignno{
r_{\Sigma N}&=1+0.19(\delta/0.2)+0.13=1.31[1.27]\ \ {\rm for}
\ \delta=0.2[0.15],\cr
r_{\Xi N}&=1-0.46(\delta/0.2)+0.43=0.97[1.09]\ \ {\rm for}
\ \delta=0.2[0.15],&(ratios)\cr
r_{\Lambda N}&=1-0.39(\delta/0.2)+0.26=0.87[0.97]\ \ {\rm for}
\ \delta=0.2[0.15].
}$$
(The choice $\delta=0.15$ corresponds to the value reported in ref. 
[\cite{kuretal}].)
\vfill
\eject
$$
\epsfbox{fig3.ps}
$$\nobreak
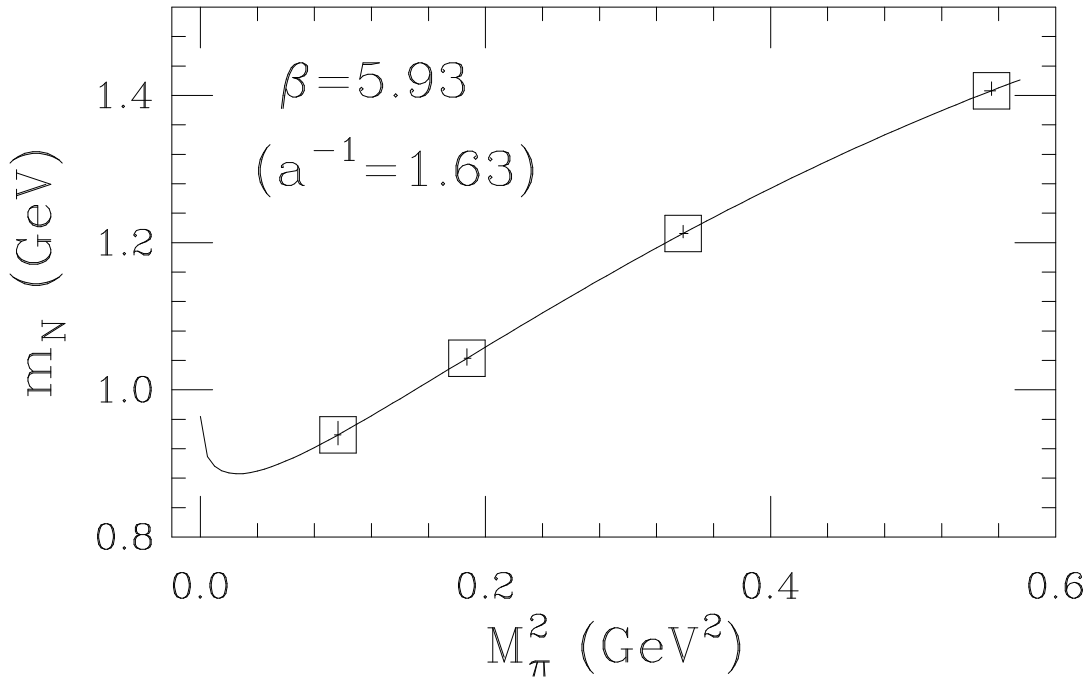
Figure 3. {\it The nucleon mass from the lattice [\cite{weinetal}] (copied
from ref. [\cite{labsha}]). The curve is from a fit to the form
$m_N=\mbar+am_\pi+bm_\pi^2+cm_\pi^3$.} \medskip
From this, one would conclude that one can expect errors from quenching 
of at least $20\%$ in the octet splittings.  These differences 
between the quenched and full theories cannot be compensated for by a 
change in scale between quenched and full QCD.

At this point I would like to comment on the above mentioned assumption
that was used in order to obtain \eq{ratios}. 
Let us consider in particular the parameters $b_D$ and $b_F$.
They correspond to higher derivative
terms in the baryon-meson effective action, and are needed in order to
absorb the UV divergences which arise at one loop in ChPT. Since the
size of these divergences is in principle different between the
full and quenched theories, one expects that $b_{D,\rm quenched}$ and
$b_{F,\rm quenched}$ can be different from $b_{D,\rm full}$ and
$b_{F,\rm full}$. 
If we want to
proceed without assuming that the quenched and full $b$'s are equal, 
we have to consider ratios of quantities
independent of the parameters $b_D$ and $b_F$. The situation is
essentially the same as in the case of $f_K/f_\pi$. From the available
results [\cite{labsha}], only one ratio independent of $b_D$ and $b_F$
can be formed:
$$X={{m_\Sigma m_\Lambda^3}\over{m_N^2 m_\Xi^2}}.\eqno(X)$$
If we expand $X$ in the Goldstone meson masses using ChPT, $X-1$
measures the deviation from the Gell-Mann--Okubo formula (\cf\ ref.
[\cite{berkaimei}] for the full theory).  

Setting $m_\pi=0$ keeping only $m_K$ as in ref. [\cite{labsha}], one 
finds
$$\eqalignno{
X_{\rm quenched}&=1-1.1046{{m_K^3}\over{8\pi \mbar 
f_\pi^2}}\left(D^2-3F^2\right)+1.3333\delta{{\pi 
m_K}\over{\sqrt{2}\mbar}}\left(D^2-3F^2\right),\cr
X_{\rm full}&=1-0.4125{{m_K^3}\over{8\pi \mbar 
f_\pi^2}}\left(D^2-3F^2\right).&(Xresults)
}$$
(The parameter $\gamma$ drops out of this particular combination and we 
again take $\alpha=0$.)
These quantities still depend on the other bare parameters, $D$, $F$ and 
$\mbar$.  Again, they could be different in the quenched and full 
theories, and I will leave the quenched values as free parameters.
Substituting $m_K=495$ $MeV$, $f_\pi=132$ $MeV$, $D_{\rm full}=0.75$, 
$F_{\rm full}=0.5$ and $\mbar_{\rm full}=1$ $GeV$ [\cite{berkaimei}]
finally gives
$${{X_{\rm quenched}}\over{X_{\rm full}}}=
1-0.0214
+\left[0.293{{\delta}\over{0.2}}-0.306\right]
{{\left(D^2-3F^2\right)_{\rm quenched}}\over{\mbar_{\rm quenched}/
1\ GeV}}.\eqno(Xratio)$$
For any reasonable values of $\mbar$ and $\delta$, and for 
$\left(D^2-3F^2\right)_{\rm quenched}$ not too far from its full theory 
value of $-0.1875$, the difference 
between the quenched and full theories as measured by the ratio $X_{\rm 
quenched}/X_{\rm full}$ is not more than a few percent.  Of course the 
same comment that applied in the case of $f_K/f_\pi$ applies here, that 
part of the difference may have been
 washed out by taking ``ratios of ratios".  
Summarizing, the conclusion of this analysis seems to be that the error 
from quenching for octet baryon masses is at least a few percent, and 
could be as much as $20\%$.

\subhead{\bf 4. A sickness of quenched QCD}

Let us again consider the quenched result for 
$f_K/f_\pi$, \eq{fkfpi}, as a
function of the quark masses (using treelevel relations between meson
masses and quark masses),
$$\left({{f_K}\over{f_\pi}}\right)^{\rm 1-loop}_{\rm quenched}=
1+{{\mu^3/3}\over{16\pi^2f_\pi^2}}
\left[{{m_u+m_s}\over{2(m_s-m_u)}}\log{{m_s}\over{m_u}}-1\right]
+{\rm (L-terms)}.$$
From this expression it is clear that we cannot take $m_u\to 0$ keeping
$m_s$ fixed, or, to put it differently, that if we take both $m_u$ and
$m_s$ to zero keeping the ratio fixed, the limit depends on this ratio,
and is not equal to one! This is quite unlike the case of full ChPT,
where one can take any quark mass to zero uniformly, and deviations
from $SU(3)$ symmetry due
to this quark mass vanish in this limit. Technically, the reason for this
strange behavior is that there is another mass $\mu$, which, as we
argued before, cannot be avoided in quenched ChPT. This mass is related 
to the singlet part of the $\eta'$ mass, and is not a free parameter of 
the theory.  Even if we do not consider any Green's functions with 
$\eta'$ external lines, this mass shows up through the double 
pole term in \eq{prop} on internal lines.  Because of the double pole, 
such contributions can lead to new infrared divergences in the $m_\pi\to 
0$ limit.  This problem with the chiral limit of quenched ChPT shows up 
in other quantities, such as meson masses and $\langle\psib\psi\rangle$ 
[\cite{us1,sharpe2,mghd}].

A question one might ask is whether this problem is an artifact of
one loop quenched ChPT [\cite{uslat92}]. For instance, if we would sum
all contributions to the $\eta'$ propagator, maybe the double
pole term would become softer in the $p\to 0$ limit, improving the
infrared behavior of diagrams in which the double pole terms appear. Let
us address this question in the chiral limit, $m_a=0$, where the problem
is most severe. In the full theory, we can write the fully dressed
$\eta'$ propagator as
$${{Z(p)}\over{p^2+\Sigma(p)}},\eqno(fulleta)$$
and define $\mu^2_F(p)=\Sigma(p)$, which onshell is the $\eta'$ mass in 
the chiral limit.  Likewise, in the quenched theory we can write the 
dressed $\eta'$, $\etat'$ propagator as
$$Z_Q(p)\left[{1\over{p^2}}\left(\matrix{1&0\cr 0&-1\cr}\right)
-{{\mu^2_Q(p)}\over{(p^2)^2}}\left(\matrix{1&1\cr 1&1\cr}\right)\right],
\eqno(quenchedeta)$$
which defines $\mu^2_Q(p)$.  To leading order in $1/N_c$, these two 
definitions of $\mu^2(p)$ should lead to the same result:
$$\mu^2_Q(p)=\mu^2_F(p)\left(1+O\left({1\over{N_c}}\right)\right).$$
We also believe that $\mu^2_F(p=0)$ is not equal to zero, since we 
expect the $\eta'$ to remain a well-behaved, 
massive particle in the chiral limit.  This 
implies, in sofar as we can rely on the large $N_c$ expansion, that 
$\mu^2_Q(0)\ne 0$, and that the double pole in \eq{prop} is a true 
feature of the theory.

While this argument is not very rigorous, I believe that the foregoing
discussion implies that the chiral limit of quenched QCD really does not
exist.  This believe is futhermore supported by the following remarks:

$\bullet$ Sharpe [\cite{sharpe2}] has summed a class of diagrams in the 
case of degenerate quark masses for a very simple quantity (the pion 
mass), and found a result that is actually more divergent than the one 
loop result.

$\bullet$ With nondegenerate quark masses there are many more diagrams 
that are infrared divergent in the chiral limit, and it is even less 
probable that resummation will improve the situation.

$\bullet$ Any mechanism improving the infrared behavior would have to 
work for each divergent quantity.  One expects that such a mechanism 
would be related to the double pole term in the $\eta'$ propagator, which 
created the problem in the first place.  But this seems unlikely in view 
of the arguments given above.

$\bullet$ The bare quark mass parameter appearing in the chiral 
effective action is not the same as that appearing in the 
(unrenormalized) QCD lagrangian.  But one can argue that the two bare 
quark masses should be analytically related, and the infrared problem is 
not just a problem of quenched ChPT, but of quenched QCD.

\subhead{\bf 5. Conclusion}

The \qa\ leads to an unknown systematic error in all lattice
calculations that employ this approximation. It would of course be very
nice to have a parameter that interpolates between full and quenched
QCD, and in principle the quark masses could play such a role, since one
expects that quenched QCD corresponds to full QCD with very heavy
quarks. One would have to distinguish here between valence and sea quark
masses, and it is the sea quark mass that would play the role of such a
parameter. This distinction can indeed be made by considering so-called
partially quenched theories [\cite{us2}], but no practical scheme to
implement this idea is known.

Quenched QCD can be defined from a euclidean pathintegral as rigorously
as full QCD. In this talk I have explained that euclidean quenched ChPT
can be used as a tool for a systematic investigation of quenched QCD.
Quenched ChPT does not quite accomplish a task equivalent to that of an
interpolating parameter. Since the bare parameters appearing in the
quenched and full chiral effective actions are not the same (as
explained in section 3) one cannot directly compare quantities
calculated in full and in quenched ChPT. However, one can calculate
combinations of physical quantities which do not depend on the bare
parameters, and in that case a direct comparison between quenched and
full QCD is possible, as we demonstrated with an example involving meson
decay constants. This makes it possible to estimate lower bounds on the
differences which come from quenching; these estimates are dependent on
the values of the meson masses, which can be taken to be the (known)
independent parameters of the theory. For realistic values of these
masses, such differences turn out to be of the order of a few percent
for ratios of decay constants and for baryon octet splittings.

The disadvantage of this more conservative approach is that part of the
difference maybe hidden, because these specific combinations of physical
quantities maybe less sensitive to the effects of quenching
than other quantities of interest. This
is particularly clear in the example of baryon octet masses. In this
case, a comparison based on the assumption that the bare parameters of
the full and quenched effective theories are the same, lead to
differences of up to $20\%$ and more.  Of course, it is not known to 
what extend this assumption is valid.

The differences between the quenched and full theories become markedly
larger for decreasing quark masses. This is due to the fact that new
infrared divergences occur in quenched QCD, which do not have a
counterpart in full QCD. These divergences lead to the nonexistence of
the chiral limit for quenched ChPT (as discussed in section 4). The
origin of this phenomenon can be traced to the special role of the
$\eta'$ in the \qa. In the \qa, the $\eta'$ is a Goldstone boson (it
develops massless poles in the chiral limit), but an additional double
pole term arises in its propagator, rendering it a ``sick" particle. For
nondegenerate quark masses this problem is also inherited by the $\pi^0$
and the $\eta$. In section 4 I argued that the nonexistence of the
chiral limit is a fundamental feature of quenched QCD.

In principle therefore, the chiral expansion breaks down for quenched
QCD. For very small quark masses, at fixed values of the singlet part of
the $\eta'$ mass $\mu^2$, the expansion becomes unreliable. In order to
make progress, one may take the expansion to be an expansion in
$\mu^2/3$ (which was shown to be roughly equal to $m_K^2$
phenomenologically), with coefficients which are functions of the quark
mass. These functions then can be expanded in terms of the quark masses,
sometimes leading to divergent behavior of the leading term (\eg\ the
one loop correction to $f_K/f_\pi$). If such divergences occur, the
expansion is only valid for a range of quark masses which are neither
too small, nor too large. It would be interesting to see whether this
point of view can be made solid.

It is in principle interesting to study any quantity which is being
computed in quenched lattice QCD in ChPT, for those quantities for which
ChPT is applicable (meson masses, decay constants, condensates and the
kaon B parameter have been calculated [\cite{us1, sharpe1,uslat92}]). As
discussed, this includes not only Goldstone meson physics {\it per se},
but also chiral corrections to baryon masses [\cite{labsha}], and for
the same reason, to mesons containing heavy quarks.

Recently, also attempts have been made to compute pion and nucleon
scattering lengths [\cite{sharpe3,kuretal2}] from quenched lattice QCD.
If one tries to calculate the $I=0$ pion scattering amplitude in
quenched ChPT, one actually finds that the imaginary part is divergent
at threshold, even for nonvanishing pion mass [\cite{pionscat}]! Again,
this can be related to double pole terms in the $\eta'$ propagator. An
additional reason is that apparently euclidean quenched correlation
functions in general cannot be analytically continued to Minkowski
space-time. (The euclidean four pion correlation functions are well
defined.) Further work is needed on pion scattering lengths.

\subhead{\bf Acknowledgments} 

First, I would like to thank Claude Bernard for a
pleasant collaboration, and for very many discussions. I also thank
Steve Sharpe, Jim Labrenz, Akira Ukawa, Hari Dass, Julius Kuti and Don
Weingarten for discussions. I am grateful for the opportunity to present
this talk given to me by the organizers of WHEPP III, Madras, India,
1994. Finally, I thank the Institute of Mathematical Sciences in Madras,
and in particular Hari Dass, for hospitality. This work is supported in
part by the Department of Energy under contract 
\#DOE-2FG02-91ER40628.

\references

\refis{sharpe}
S.R.~Sharpe, in {\it The Fermilab Meeting}, 7th meeting of the DPF, eds.
C.H.~Albright, P.H.~Kasper, R.~Raja and J.~Yoh, World Scientific, 1993.

\refis{mackenzie}
P.B.~Mackenzie, preprint FERMILAB-CONF-93-343-T, to be published in the 
proceedings of the 16th International Symposium on Lepton and Photon 
Interactions, 1993, hep-ph/9311242; A.S.~Kronfeld and P.B.~Mackenzie,
Ann. Rev. Nucl. Part. Phys. {\bf 43} (1993) 793.

\refis{lat93} 
Lattice'93, proceedings of the International Symposium 
on Lattice Field Theory, 
Dallas, Texas, 1993.

\refis{quenched}
H.~Hamber and G.~Parisi, \prl{47} (1981) 1792;
E.~Marinari, G.~Parisi and C.~Rebbi, \prl{47} (1981)
1795; D.H.~Weingarten, \plb{109} (1982) 57.

\refis{sharpe1}
S.R.~Sharpe, \prd{41} (1990) 3233; Nucl. Phys. {\bf B} 
(Proc.Suppl.) {\bf 17} (1990) 146;
G.~Kilcup \etal, \prl{64} (1990) 25; S.R.~Sharpe, DOE/ER/40614-5, to be
published in {\it Standard Model, Hadron Phenomenology and Weak Decays
on the Lattice}, ed. G.~Martinelli, World Scientific.

\refis{us1}
C.W.~Bernard and M.F.L.~Golterman, \prd{46} (1992) 853; Nucl. Phys. {\bf 
B} (Proc.Suppl.) {\bf 26} (1992) 360.

\refis{sharpe2}
S.R.~Sharpe, \prd{46} (1992) 3146; Nucl. Phys. {\bf B}(Proc.Suppl.) {\bf 
30} (1993) 213.

\refis{uslat92}
C.W.~Bernard and M.F.L.~Golterman, Nucl. Phys. {\bf B}(Proc.Suppl.) {\bf 30} 
(1993) 217.

\refis{hasleu}
J.~Gasser and H.~Leutwyler, \plb{184} (1987) 83,\plb{188} (1987) 477,
and \npb{307} (1988) 763; H.~Leutwyler, Nucl. Phys. {\bf B} (Proc.Suppl.)
{\bf 4} (1988) 248
and \plb{189} (1987) 197; H.~Neuberger, \npb{300} (1988) 180;
P.~Hasenfratz and H.~Leutwyler, \npb{343} (1990) 241.

\refis{labsha}
J.N.~Labrenz and S.R.~Sharpe, preprint UW-PT-93-07, to be published 
in the proceedings of the International Symposium on Lattice Field 
Theory, Dallas, Texas, 1993, hep-lat/9312067.

\refis{morel}
A.~Morel, J. Physique {\bf 48} (1987) 111.

\refis{dewitt}
For a description of the
properties of graded groups, see for example, B.~DeWitt,
{\it Supermanifolds}, Cambridge, 1984.

\refis{gasleu}
J.~Gasser and H.~Leutwyler, \npb{250} (1985) 465.

\refis{weinberg}
S.~Weinberg, Physica {\bf 96A} (1979) 327.

\refis{kuretal}
Y.~Kuramashi, M.~Fukugita, H.~Mino, M.~Okawa and A.~Ukawa, preprint 
KEK-CP-13, 1994; preprint KEK-CP-010, to be published
in the proceedings of the International Symposium on Lattice Field
Theory, Dallas, Texas, 1993, hep-lat/9312016.

\refis{witven}
E.~Witten, \npb{156} (1979) 269;
G.~Veneziano, \npb{159} (1979) 213.

\refis{hatsuda}
T.~Hatsuda, \npb{329} (1990) 376; S.R.~Sharpe, private communication.

\refis{manjen}
E.~Jenkins and A.~Manohar, \plb{255} (1991) 558; E.~Jenkins, \npb{368} 
(1992) 190.

\refis{berkaimei}
V.~Bernard, N.~Kaiser and U.~Meissner, Z. Phys. {\bf C60} (1993) 111.

\refis{weinetal}
F.~Butler, H.~Chen, J.~Sexton, A.~Vaccarino and D.~Weingarten,
\prl{70} (1993) 2849; Nucl. Phys. {\bf B} (Proc.Suppl) {\bf 30} (1993) 377.

\refis{mghd}
M.F.L.~Golterman and N.D.~Hari~Dass, in preparation.

\refis{us2}
C.W.~Bernard and M.F.L.~Golterman, \prd{49} (1994) 486;
preprint WASH-U-HEP-93-61, to be published
in the proceedings of the International Symposium on Lattice Field
Theory, Dallas, Texas, 1993, hep-lat/9311070.

\refis{sharpe3}
R.~Gupta, A.~Patel and S.R.~Sharpe, 
\prd{48} (1993) 388.

\refis{kuretal2}
Y.~Kuramashi, M.~Fukugita, H.~Mino, M.~Okawa and A.~Ukawa,
\prl{71} (1993) 2387.

\refis{pionscat}
C.W.~Bernard, M.F.L.~Golterman, J.N.~Labrenz, S.R.~Sharpe and A.~Ukawa, 
preprint Wash.~U. HEP/93-62, to be published
in the proceedings of the International Symposium on Lattice Field
Theory, Dallas, Texas, 1993.

\endreferences 

\vfill 
\bye